\documentclass[letterpaper, 10 pt, conference]{ieeeconf}  

\IEEEoverridecommandlockouts                              
\usepackage{acronym}
\usepackage{amstext}
\usepackage{url}
\usepackage{graphicx}
\usepackage{epstopdf}
\usepackage{amsmath}
\usepackage{subfigure}
\usepackage{amsfonts}
\usepackage{algorithmic}
\usepackage{algorithm}
\usepackage{cite}
\usepackage{lipsum}
\usepackage{amssymb}

\usepackage{acronym}

\newtheorem{theorem}{Theorem}

\newtheorem{lemma}{Lemma}
\newtheorem{definition}{Definition}

\newtheorem{remark}{Remark}

\acrodef{TCL}{Thermostatically Controlled Load}
\acrodef{AC}{Air Conditioning}
\acrodef{FERC}{Federal Energy Regulatory Commission}
\acrodef{AGC}{Automatic Generation Control }
\acrodef{CAISO}{California Independent System Operator}
\acrodef{PJM}{Pennsylvania-New Jersey-Maryland}
\acrodef{MCP}{Market Clearing Price}
\acrodef{NGR}{Non-Generator Resource}
\acrodef{SoC}{State of Charge}
\acrodef{EV}{Electric Vehicle}
\acrodef{RPS}{Renewable Portfolio Standard}
\acrodef{CAISO}{California Independent System Operator}

\newcommand{\cut}[1]{}

\usepackage{tikz,pgfplots}
\usepackage[hang,flushmargin]{footmisc}

\allowdisplaybreaks

\title{Improved Battery Models of an Aggregation of \\ Thermostatically Controlled Loads for Frequency Regulatio$\text{n}^{\pi}$
}

\author{
Borhan M. Sanandaj$\text{i}^{\text{b},\star}$, He Ha$\text{o}^{\text{b}}$, 
Kameshwar Pooll$\text{a}^{\text{b}}$, and Tyrone L. Vincen$\text{t}^{\text{c}}$
\thanks{\textsuperscript{$\text{b}$}Borhan M. Sanandaji, He Hao, and Kameshwar Poolla are with the Department of Electrical Engineering and Computer Sciences, University of California, Berkeley, CA 94720.}
\thanks{\textsuperscript{$\star$}Corresponding author. Email: sanandaji@berkeley.edu.}
\thanks{\textsuperscript{$\text{c}$}Tyrone L. Vincent is with the Department of Electrical Engineering and Computer Science, Colorado School of Mines, Golden, CO 80401.}
\thanks{\textsuperscript{$\pi$}Supported in part by EPRI and CERTS under sub-award 09-206; PSERC S-52; NSF under Grants CNS-0931748, EECS-1129061, CPS-1239178, and CNS-1239274; the Republic of Singapore National Research Foundation through a grant to the Berkeley Education Alliance for Research in Singapore for the SinBerBEST Program; Robert Bosch LLC through its Bosch Energy Research Network funding program. 
}
}

\begin{document}

\maketitle

\begin{abstract}
Recently it has been shown that an aggregation of Thermostatically Controlled Loads (TCLs) can be utilized to provide fast regulating reserve service for power grids and the behavior of the aggregation can be captured by a stochastic battery with dissipation.
In this paper, we address two practical issues associated with the proposed battery model. First, we address clustering of a heterogeneous collection and show that by finding the optimal dissipation parameter for a given collection, one can divide these units into few clusters and improve the overall battery model.
Second, we analytically characterize the impact of imposing a no-short-cycling requirement on TCLs as constraints on the ramping rate of the regulation signal. We support our theorems by providing simulation results.
\end{abstract}

\acresetall


\section{Introduction}
\label{sec:intro}

\subsection{Renewable Integration and Regulating Reserve Service}
Vast and deep integration of renewable energy resources into the existing power grid is essential in achieving the envisioned sustainable energy future. Environmental, economical, and geopolitical concerns associated with the current power grid have motivated many countries around the globe as well as many states in the U.S. to setup aggressive \acp{RPS}. The state of California, as an example, has targeted a $33\%$ \ac{RPS} by 2020 \cite{CA_renewable_portfolio}. 
Volatility, stochasticity, and intermittency characteristics of renewable energies, however, present challenges for integrating these resources into the existing grid in a large scale as the proper functioning of an electric grid requires a continuous power balance between supply and demand. 

Ancillary services such as regulating reserve (or frequency regulation) and load following play an important role in maintaining a functional and reliable grid under normal conditions~\cite{smith2007utility, makarov2009operational, meynegwankowsha10}.~While load following handles more predictable and slower changes in load, regulating reserve handles imbalances at faster time scales~\cite{AS_Kirby}.
On the other hand, an increased penetration of renewable energies results in higher regulation requirements on the grid \cite{smith2007utility, makarov2009operational, meynegwankowsha10}. For instance, it has been shown that if California adopts its $33\%$ \ac{RPS} by 2020, the regulation procurement is anticipated to increase from $0.6$ GW to $1.4$ GW \cite{helman2010resource,CAISO_flexible}. Such requirements can be lowered if \emph{faster} responding resources are available~\cite{kema2009fastresponse}. For instance, it has been shown that if \ac{CAISO} dispatches fast responding regulation resources, it would reduce its regulation procurement by $40\%$~\cite{pnnl2008value}.

\subsection{Demand-Side Flexibility for Frequency Regulation}
Frequency regulation is one of the most important ancillary services for maintaining the power balance in normal conditions~\cite{AS_Kirby}. It is deployed in seconds (up to one minute) time scales to compensate for short term fluctuations in the \emph{net load}.\footnote{Net load is defined as forecasted load minus predicted variable generation.} This service has been traditionally provided by either fast responding generators or grid-scale energy storage units. However, the current storage technologies such as batteries have high cost while generation has both cost and an environmental footprint. Moreover, traditional generators have slow ramping rates and cannot track the fast changing regulation signal very well. These factors coupled with the search for cleaner sources of flexibility as well as regulatory developments such as \ac{FERC} order 755 (2011)~\cite{ferc_755} and 784 (2013)~\cite{ferc_784} have motivated a growing interest in tapping fast responding demand-side resources for enabling deep renewable integration. 

%

\subsection{Aggregation of Flexible Loads} 
Flexible loads such as \acp{EV} and residential and commercial buildings have been recently considered as good candidates for providing ancillary services to the grid~\cite{callaway2009tapping,koch2009active,koch2011modeling,mathieu2012state,mathieu2013state,mathieu2013energy,zhang2012aggregate,chang2013modeling,zhang2013aggregated,bashash2011modeling,bashash2013modeling,sanandaji2014fast,bashash2011robust,nayyar2013EV,maasoumy2014flexibility}. 
Residential \acp{TCL} such as air conditioners, heat pumps, water heaters, and refrigerators, represent about $20\%$ of the total electricity consumption in the United States \cite{building_energy_data_book,eia_aer}, and thus present a large potential for providing various ancillary services to the grid. Leveraging the inherent thermal slackness of \acp{TCL}, their electricity consumption can be varied while still meeting the desired comfort level and temperature requirements of the end user.  
\subsection{Related Work}

\acp{TCL} have been recently considered for providing load following and regulation services to the grid~\cite{callaway2009tapping,koch2009active,koch2011modeling,kundu2011modeling,mathieu2012state,mathieu2013energy}.
In particular, it has been recently shown that the aggregate flexibility offered by a collection of \acp{TCL} can be succinctly modeled as a {\em stochastic battery} with dissipation~\cite{hehao2013generalized,hehao2013aggregate}. 
The power limits and energy capacity of this battery model can be calculated in terms of \ac{TCL} model parameters and exogenous variables such as ambient temperature and user-specified set-points.
Simple battery models are also considered in~\cite{koch2009active,mathieu2013energy}. Clustering and no-short-cycling of \acp{TCL} have been 
reported in~\cite{zhang2012aggregate,chang2013modeling}. 
\subsection{Main Contributions}
In this paper, we address some practical aspects associated with our earlier proposed battery model\cite{hehao2013generalized,hehao2013aggregate}.
First, we consider the impact of dividing a heterogeneous collection of \acp{TCL} into clusters and show that by finding the optimal dissipation parameter for a given collection of \acp{TCL}, one can divide these units into a few stochastic batteries and improve the battery model. 
Second, we consider the effect of enforcing a requirement of no-short-cycling. In order to avoid damages, \acp{TCL} manufacturers require a minimum duration of time between any two switches of ON/OFF state. If this minimum time is not met, the unit is said to be short-cycled.
In particular, we show that the no-short-cycling constraint can be expressed as constraints on the ramping rate (first difference) of the~\ac{AGC} signal. 
Consequently, a characterization of regulation signals that can be feasibly met by a \ac{TCL} aggregation is the intersection of signals feasible for the stochastic battery model, and this new constraint on the first difference of the regulation signal. To the best of our knowledge, this work is the first which explicitly represents a ramping rate constraint on the regulation signal as a consequence of units no-short-cycling requirements.
%



\subsection{Paper Organization}
The remainder of the paper is organized as follows. Section \ref{sec:model} describes preliminaries on individual \ac{TCL} models. In Section~\ref{sec:battery}, we summarize the stochastic battery model. Optimal dissipation and clustering of a collection of \acp{TCL} are presented in Section~\ref{sec:clustering}. We address the no-short-cycling of \acp{TCL} in Section~\ref{sec:short-cyling}. Whenever needed and within each section, we provide simulation results to support our theorems. 

\section{Thermostatically Controlled Loads}
\label{sec:model}
The temperature evolution of the $k$th \ac{TCL} can be described by a standard dead-band model as
\begin{align}
 \dot{\theta}^k(t) =
  \begin{cases} 
	 -a^k(\theta^k(t)-\theta_a) - b^kP^k_m + w^k(t), & \text{ON state},\\	
	 -a^k(\theta^k(t)-\theta_a)  + w^k(t), & \text{OFF state},
\end{cases}
\label{eq:hybrid_model}
\end{align}
where $\theta^k(t)$ is the internal temperature of the $k$th \ac{TCL} at time $t$, $\theta_a$ is the ambient temperature, $P^k_m$ is the rated electrical power, $a^k := 1/C^k R^k, b^k := \eta^k/C^k$, and $R^k$, $C^k$, and $\eta^k$ are model parameters as described in Table~\ref{tab:model_parameters}. For more details on the \ac{TCL} model, please see~\cite{callaway2009tapping, koch2011modeling, hehao2013generalized}.\footnote{
Four types of \acp{TCL} are: (i) air conditioners, (ii) heat pumps, (iii) water heaters, and (iv) refrigerators.}
Each \ac{TCL} has a temperature set-point $\theta^k_r$ with a hysteretic ON/OFF local control within a dead-band $[\theta^k_r - \Delta^k, \theta^k_r + \Delta^k]$. The operating state of the $k$th \ac{TCL}, $q^k(t)$, evolves as
\begin{align*}
\lim_{\epsilon \rightarrow 0} q^k(t+\epsilon) =
\begin{cases} 
	q^k(t), & |\theta^k(t) - \theta^k_r| < \Delta^k, \\
	1-q^k(t), & |\theta^k(t) - \theta^k_r| = \Delta^k,
\end{cases}
\end{align*}
where $q^k(t) = 1$ when the \ac{TCL} is ON and $q^k(t) = 0$ when it is OFF. 
The average power consumed by the $k$th \ac{TCL} over a cycle is 
\begin{align*}
\scriptsize
	P^k_a =  \frac{P^k_m T^k_{\textrm{ON}}}{T^k_{\textrm{ON}}+T^k_{\textrm{OFF}}},
\end{align*}
where $T^k_{\textrm{ON}}$ and $T^k_{\textrm{OFF}}$ are given by
\begin{align*}
	T^k_{\textrm{ON}} &= R^kC^k \text{ ln } \frac{\theta^k_r + \Delta^k -\theta_a+R^kP^k_m\eta^k}{\theta^k_r - \Delta^k-\theta_a+R^kP^k_m\eta^k}, \\
	T^k_{\textrm{OFF}} &= R^kC^k \text{ ln } \frac{\theta^k_r - \Delta^k-\theta_a}{\theta^k_r + \Delta^k -\theta_a},
\end{align*}
and represent the ON and OFF state durations per cycle, respectively. 
For a large collection of \acp{TCL} that is uncoordinated, the instantaneous power drawn by this collection will be very close to the combined average power requirement due to the fact that any specific \ac{TCL} will be at a uniformly random point along its operating cycle. 
For a heterogeneous collection of \acp{TCL} indexed by $k$, the baseline power is 
\begin{align*}
\scriptsize
	n(t) := \sum_{k}  P^k_a.
\end{align*}
The aggregated instantaneous power consumption is 
\begin{align*}
\scriptsize
	P_{\textrm{agg}}(t) := \sum_{k}  q^k(t) P^k_m.
\end{align*}

%


\begin{table}[tb]
\caption{Typical parameter values for a residential Air Conditioner.}
\vspace{-1.5em}
\label{tab:model_parameters}
\begin{center}
\begin{tabular}{cccc}
\hline
Parameter & Description & Value & Unit\\
\hline 
$C$ & thermal capacitance & $2$ & kWh/$^{\circ}{\rm C}$\\
$R$ & thermal resistance &  $2$ & $^{\circ}{\rm C}$/kW\\
$P_m$ & rated electrical power & $5.6$ & kW\\
$\eta$ & coefficient of performance& $2.5$ & \\
$\theta_r$ & temperature set-point & $22.5$ &$^{\circ}{\rm C}$\\
$\Delta$ & temperature deadband & 0.3125 & $^{\circ}{\rm C}$\\
$\theta_a$ & ambient temperature & $32$ & $^{\circ}{\rm C}$\\
\hline
\end{tabular}
\vspace{-1.5em}
\end{center}
\end{table}

As an approximation to the dead-band model, we consider a continuous-power model. Here, a \ac{TCL} accepts any continuous power input $p^k(t) \in [0, P^k_m]$ and the dynamics are
\begin{align*}
\dot{\theta}^k(t) = -a^k(\theta^k(t)-\theta_a) - b^k p^k(t).
\label{eq:continuous_model}
\end{align*}
As common in the literature, the disturbance $w^k(t)$ in Model \eqref{eq:hybrid_model} is assumed to be Gaussian with zero mean and small variance \cite{malhame1985electric,callaway2009tapping,
koch2011modeling}, and can be neglected.
Maintaining the temperature $\theta^k(t)$ within the user-specified dead-band $\theta^k_r \pm \Delta^k$ is treated implicitly as a {\em constraint} on the power signal $p^k(t)$. When evaluating the trajectory $\theta^k(t)$, it is assumed that $\theta^k(0)=\theta^k_r$. The parameters that specify this continuous-power model are $\chi^k = (a^k,b^k, \theta^k_r, \theta_a, \Delta^k, P^k_m)$.
The nominal power required to keep the $k$th \ac{TCL} at its set-point is 
\begin{align*}
\scriptsize
P^k_o = \frac{a^k(\theta_a -\theta^k_r)}{b^k}=\frac{\theta_a -\theta^k_r}{\eta^k R^k}.
\end{align*}

We note that $P^k_o$ is a random process as it depends on the ambient temperature and the user-defined set-point. Simple calculations with typical parameters reveal that the nominal power $P^k_o$ under the continuous-power model closely follows the average power $P^k_a$ under the dead-band model for a wide range of operating conditions. In \cite{sanandaji2014fast,hehao2013generalized}, we showed that the aggregate behavior of a population of \acp{TCL} with the dead-band model could be accurately approximated by the those using the continuous-power model. The continuous-power model is used for analysis, and the dead-band model is used for simulations.

\section{Stochastic Battery Model}
\label{sec:battery}

Each \ac{TCL} can accept perturbations around its nominal power consumption ($p^k(t) = P_o^k+e^k(t)$) that will meet user-specified comfort bounds.
Define 
\[ \mathbb{E}^k := \left\{ e^k(t) \Big | \begin{matrix}  0 \leq P_{o}^{k} + e^{k}(t)  \leq P_{m}^{k}, \\ P_o^k +e^k(t) \text{ maintains } |\theta^k(t) - \theta_r^k| \leq \Delta^k
 \end{matrix} \right\}. 
 \]
This set of power signals represents the flexibility of the $k$th \ac{TCL} with respect to its nominal.
The \emph{aggregate flexibility} of the collection of TCLs is defined as the Minkowski sum
\begin{align*}
\mathbb{U} = \sum_k \mathbb{E}^k.
\end{align*}

The difficulty is in evaluating $\mathbb{U}$ as its geometry is, in general, bulky. In \cite{hehao2013generalized,hehao2013aggregate}, we showed that the set $\mathbb{U}$ can be nested within two generalized battery models. 

\begin{definition}
Let $\phi = (\mathcal{C}, n_-, n_+, \alpha)$ be non-negative parameters. A \emph{Generalized Battery Model}, $\mathbb{B}(\phi)$, is a set of signals $u(t)$ that satisfy
\begin{align*}
-n_- \leq u(t) \leq n_+, \quad \forall \ t>0, \qquad \qquad \\
\dot{x}(t) = -\alpha x(t)- u(t), \ x(0) = 0 \  \Rightarrow |x(t)| \leq \mathcal{C}, \quad \forall \ t>0.
\end{align*} 
\label{def:battery}
\end{definition}

One can regard $u(t)$ as the power drawn from or supplied to a battery and $x(t)$ as its \ac{SoC}.~One should note that the parameters $\phi$ are random and depend on ambient temperature and participation rates. As a result, we regard $\mathbb{B}(\phi)$ as a stochastic battery. This battery model provides a compact framework to characterize the aggregate power limits and energy capacity of a population of \acp{TCL}. 
\begin{theorem}[\cite{hehao2013generalized}]
Consider a heterogeneous collection of \acp{TCL} modeled by the continuous-power model with parameters $\chi^k$.
Let $\alpha>0$ be the dissipation parameter. Let
\[
f^k := \Delta^k/(b^k(1+|1-\alpha/a^k|)).
\] 
The aggregate flexibility $\mathbb{U}$ of the collection satisfies 
\begin{align*}
\mathbb{B}(\phi_s) \subseteq \mathbb{U} \subseteq \mathbb{B}(\phi_n),
\end{align*}
where the necessary battery model parameters are given by
\begin{align}
\phi_n :
\begin{cases}
\mathcal{C} = \sum_k \left(1+\left|1-\frac{a^k}{\alpha}\right|\right)\frac{\Delta^k}{b^k},\\
n_- = \sum_kP_o^k,\\
n_+ = \sum_k (P_m^k - P_o^k),
\end{cases}
\end{align}
For a given $\alpha$, the sufficient battery model parameters are any triple ($\mathcal{C}, n_-,n_+$) that satisfies  
\begin{align}
\phi_s : 
\begin{cases}
\beta^k n_- \leq  P_o^k, \ \ \ \ \ \ \ \ \; \; (\forall k)\\
\beta^k n_+ \leq P_m^k - P_o^k, \ \  (\forall k)\\
\beta^k \mathcal{C} \leq f^k, \ \ \ \ \ \ \ \ \ \ \ \ (\forall k)
\end{cases}
\label{eq:sufficient_condition}
\end{align}
where $\beta^k \geq 0$ satisfies $\sum_k \beta^k=1$. Further, if $u(t) \in \mathbb{B}(\phi_{s})$, the causal power allocation strategy
\[
e^k(t) = \beta^k u(t)
\]
satisfies the deadband constraints $|\theta^k(t) - \theta_r^k| \leq \Delta^k$.
\label{thm:main_results}
\end{theorem}

One should note that the gap between the proposed battery models $\mathbb{B}(\phi_s)$ and $\mathbb{B}(\phi_n)$ in Theorem~\ref{thm:main_results} depends on the choice of allocation $\beta^k$, the dissipation $\alpha$, and heterogeneity level of the collection of \acp{TCL}.
In the next section, we explain how we can obtain an optimal dissipation for a given collection of \acp{TCL}. Moreover, we show how one can improve the battery models by means of clustering of units. 

%

\section{Optimal Dissipation and Clustering of TCLs}
\label{sec:clustering}
As mentioned earlier, there exist different choices of $\beta_k$ that satisfy~(\ref{eq:sufficient_condition}). For each choice, a different battery model $\mathbb{B}(\phi_s)$ will be obtained that assures feasibility.
One choice is
\begin{equation}
\beta_k = \frac{P_m^k - P_o^k}{\sum_k (P_m^k - P_o^k)},
\label{eq:allocation_n_+}
\end{equation}
which yields the smallest gap between the necessary and sufficient battery models for $n^+$ as compared to other choices of $\beta^k$. However, it results in larger gaps for $n^-$ and $\mathcal{C}$.\footnote{We can maximize the bounds on $\mathcal{C}$ and $n^-$ by choosing different allocations $\beta_k$. Please refer to~\cite{hehao2013aggregate} where we discuss how different choices of $\beta_k$ would affect the bounds on power limits and energy capacity.} 
Based on this particular choice of $\beta_k$, 
\begin{align}
\phi_s:
\begin{cases}
\mathcal{C} = \sum_k (P_m^k-P_o^k)\min_k  \frac{f^k}{P_m^k-P_o^k},\\
n_- = \sum_k (P_m^k - P_o^k)\min_k \frac{P_o^k}{P_m^k - P_o^k},\\
n_+ = \sum_k (P_m^k - P_o^k).
\end{cases}
\label{eq:sufficient_condition_n_+}
\end{align}
This result is valid for any given dissipation parameter $\alpha$. 
\subsection{Optimal Dissipation Parameter}
\label{sec:optimal}
While the bound on $n_+$ is the tightest possible based on the allocation~(\ref{eq:allocation_n_+}), one would like to tighten the bound on $\mathcal{C}$ as well. This can be done easily by considering the following optimization problem for a given heterogeneous collection:
\begin{equation}
\alpha^*:= \arg \max_{\alpha} \min_k  \frac{f^k}{P_m^k-P_o^k}
\label{eq:max_min_a}
\end{equation}
to find the \emph{optimal dissipation} parameter $\alpha^*$. While solving (\ref{eq:max_min_a}) for the general case may require a numerical solution, for some heterogeneity scenarios we get an analytical solution. 
\subsubsection{Thermal Capacity}
\label{sec:thermal_capacity}

Consider the case where all of the parameters are homogenous and the only heterogeneity is in $C^k$.\footnote{In total there are $6$ parameters whose heterogeneity can affect $\mathcal{C}$ in (\ref{eq:sufficient_condition_n_+}): $C^k$, $R^k$, $\eta^k$, $\Delta^k$, $P_m^k$, and $\theta_r^k$.} 
Then, the battery capacity can be written as
\[
\mathcal{C}(\alpha) = N\Delta (\min_k g^k)/\eta,
\]
where $g^k := C^k/(1+|1-\alpha RC^k|)$. Note that the dependance of the capacity on $\alpha$ has been made explicit.
%
%
%
\begin{lemma}
Consider a heterogeneous collection of \acp{TCL} where the heterogeneity is only in $C^k$. Then
\[
\mathcal{C}^* := \max_{\alpha} \mathcal{C}(\alpha) = N\Delta C_{\min} /\eta \ \ \text{and} \ \ \alpha^* = 1/RC_{\min},
\]
where $C_{\min} := \min_k C^k$.
\label{lem:hetero_thermal}
\end{lemma}
\begin{proof}
See Appendix.
\end{proof}

\subsubsection{Dead-band}
\label{sec:deadband}

Consider the case where all of the parameters are homogenous and the only heterogeneity is in $\Delta^k$. Then, the battery capacity can be written as
\[
\mathcal{C}(\alpha) = NC(\min_k g^k)/(\eta({1+|1-\alpha RC|})),
\]
where $g^k := \Delta^k$.
\begin{lemma}
Consider a heterogeneous collection of \acp{TCL} where the heterogeneity is only in $\Delta^k$. Then
\[
\mathcal{C}^* = NC\Delta_{\min}/\eta \ \ \text{and} \ \ \alpha^* = 1/RC,
\]
where $\Delta_{\min} := \min_k \Delta^k$. 
\label{lem:hetero_deadband}
\end{lemma}
\begin{proof}
See Appendix.
\end{proof}

%
%
One can derive similar results for cases where more parameters contain heterogeneity. For example, when the heterogeneity is in both $C^k$ and $\Delta^k$, 
then $\mathcal{C}(\alpha) = N(\min_k g^k)/\eta$ 
where $g^k := \frac{C^k\Delta^k}{1+|1-\alpha RC^k|}$.
One can show that under this assumption,
\[
\mathcal{C}^* \gtrapprox NC_{\min}\Delta_{\min}/\eta
 \ \ \text{and} \ \ \alpha^* \lessapprox 1/RC_{\min}.
\]
\subsection{Optimal Clustering}

\begin{figure}[tb]
\centering
\includegraphics[width=\columnwidth]{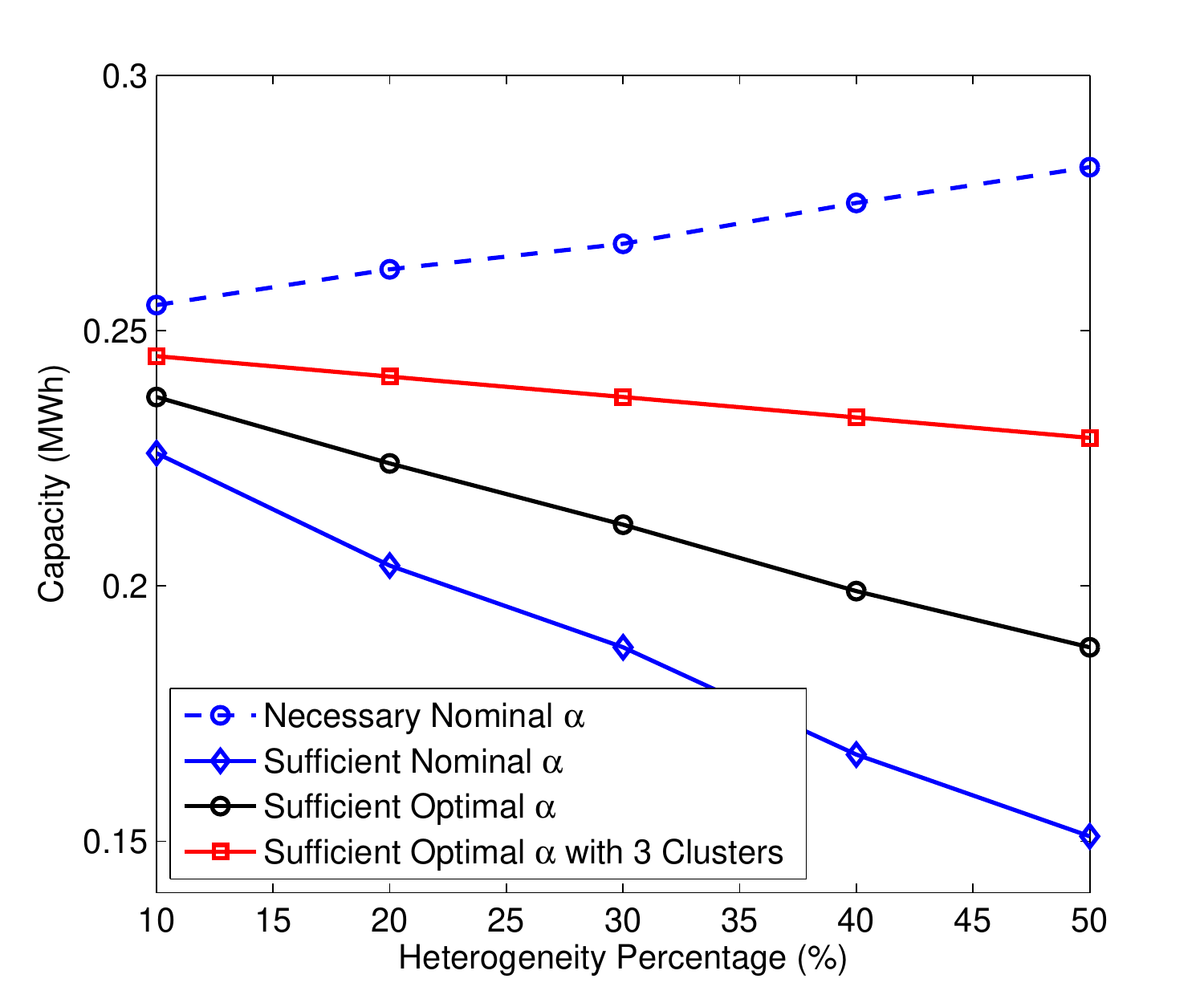}
\caption{The effect of optimal dissipation parameter and clustering on the battery model. A collection of $1000$ heterogenous \acp{TCL} are considered whose nominal parameter values are given in Table~\ref{tab:model_parameters}. A uniform distribution is considered as the heterogeneity pattern of $C^k$.}
\label{fig:clustering_bound_comparison_1}
\end{figure}

As the diversity of the \ac{TCL} model parameters increases,
the gap between $\mathbb{B}(\phi_s)$ and $\mathbb{B}(\phi_n)$ increases. 
Fig.~\ref{fig:clustering_bound_comparison_1} illustrates how the sufficient and necessary capacity values provided by $\mathbb{B}(\phi_s)$ and $\mathbb{B}(\phi_n)$, respectively, increases as the heterogeneity level increases.
In order to improve the battery models, one can divide a heterogenous collection~\acp{TCL} into a few clusters and derive battery models for each of those clusters. 
The number of clusters should be small in order to keep the benefits of aggregation within each battery model, while limiting the complexity of the overall model. We assume the number of clusters $m$ is small and given. 
 
%
%
Consider the case where only $C^k$ contains heterogeneity.
%
We know that from our previous analysis, for a given collection of $N$ \acp{TCL} with heterogeneous $C^k$'s, $N\Delta C_{\min}/\eta$ is the maximum energy capacity~(Lemma~\ref{lem:hetero_thermal}). Let's assume we want to divide $N$ units into $m$ clusters where $N_i$ is the size of cluster $i$ such that $\sum_{i=1}^mN_i = N$. The goal is to find optimal cluster sizes such that the overall energy capacity of the collection is maximized. The following theorem provides the optimal cluster sizes and optimal capacity under clustering when there is a uniform distribution on $C^k$'s. 
\begin{theorem}
Consider a heterogenous collection of $N$ \acp{TCL} with model~(\ref{eq:hybrid_model}) where $C^k$ has a uniform distribution as
$C^k \sim U(C_{\min},C_{\max})$ but all other parameters are identical between units. Then, an optimal clustering is achieved by sorting the units based on their $C^k$ value and then putting the first $N/m$ units in the first cluster, the second $N/m$ units in the second cluster, etc, with an optimal cluster size of
\begin{equation}
N^*_i = N/m, \qquad i=1,2,\dots,m,
\label{eq:optimal_cluster_size_Ck}
\end{equation}
where $m$ is the number of clusters. The optimal capacity is 
\begin{equation}
\hspace{-0.037in}\mathcal{C}^*_m = \big(C_{\min}+\frac{(C_{\max}-C_{\min})}{2}\frac{N}{(N-1)}\frac{(m-1)}{m}\big)N\Delta/\eta,
\label{eq:optimal_capacity_Ck}
\end{equation}
where $\mathcal{C}^*_m$ is the optimal capacity with $m$ clusters. 
\label{theo:clustering}
\end{theorem}
\begin{proof}
See Appendix.
\end{proof}
\begin{remark}
Similar results can be achieved when more parameters contain heterogeneity (and even with different heterogeneity distributions) following the steps explained in the proof of Theorem~\ref{theo:clustering} and based on the optimal dissipation parameters as discussed in Section~\ref{sec:optimal}.
\end{remark}

Fig.~\ref{fig:clustering_bound_comparison_1} illustrates how the sufficient and necessary capacity bounds change over different heterogeneity levels and under different scenarios on the dissipation parameter and clustering. 
As can be seen, when a nominal (an average of the time constants $1/RC^k$) dissipation is considered with no clustering, the gap between the sufficient and necessary capacity bounds increases as the heterogeneity level increases. 
This gap can be decreases when an optimal dissipation parameter is used for the collection. Moreover, the gap can be further tightened when we divide the collection into a few clusters (in this example $3$ clusters). Apparently when $m=1$, the optimal capacity is the same as provided by Lemma~\ref{lem:hetero_thermal}. A keen reader would note that when $m=N$, the optimal capacity~(\ref{eq:optimal_capacity_Ck}) is 
\[
\mathcal{C}^*_N = \frac{N\Delta C}{\eta},
\]
which is the capacity bound of a homogenous collection~\cite{hehao2013generalized,hehao2013aggregate}. As mentioned earlier, we keep $m$ small such that the complexity of the overall battery model is as low as possible.


\section{No-Short-Cycling and Ramping Rate Constraints}
\label{sec:short-cyling}
In this section, we first present our priority-stack-based control framework for manipulating the power consumption of a population of \acp{TCL} and for providing regulation service to the grid. We then augment our control structure with a no-short-cycling constraint. Moreover, we analytically characterize the no-short-cycling constraint in terms of bounds on the ramping rate of the regulation signal.  
\subsection{Priority-Stack-Based Control}
We adopt a centralized control architecture. This choice is dictated by the stringent power quality, auditing and telemetry requirements necessary to participate in regulation service market~\cite{hehao2014frequency}. At each sample time, the aggregator compares the regulation signal $r(t)$ with the aggregate power 
deviation $\delta(t) = P_{\textrm{agg}}(t)-n(t)$, where $P_{\textrm{agg}}(t)$ is the instantaneous power drawn by \acp{TCL} and $n(t)$ is their baseline power.  

If $r(t)<\delta(t)$, the population of \acp{TCL} needs to ``discharge'' power to the grid which requires turning OFF some of the ON units. Conversely, if $r(t)>\delta(t)$, then the population of \acp{TCL} must consume more power. This requires turning ON some of the OFF units. To track a regulation signal $r(t)$, the system operator needs to determine appropriate switching actions for each \ac{TCL} so that the power deviation of \acp{TCL}, $\delta(t)$, follows the regulation signal $r(t)$.

\begin{figure}[tb]
	\centering
	\includegraphics[width=.9\columnwidth]{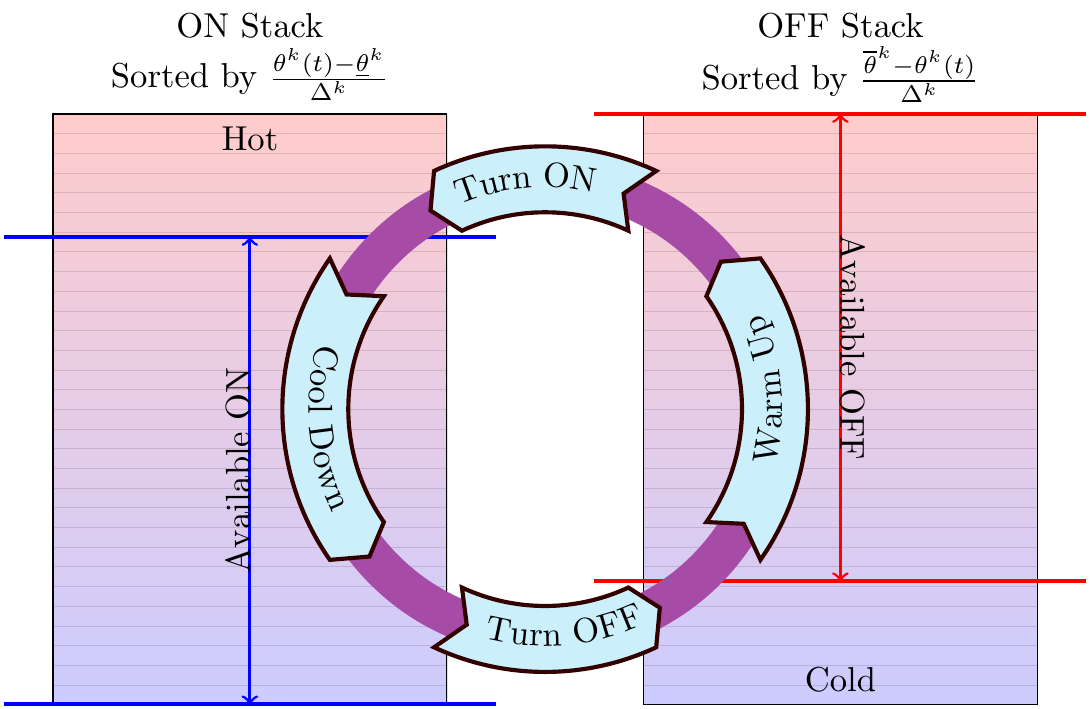}
	\caption{The ON and OFF priority stacks with 
	explicit no-short-cycling constraints. A unit that is hotter has a higher priority to be switched ON and a unit that is cooler has a higher priority to be turned OFF. However, when no-short-cycling constraints are imposed, we are only allowed to manipulate units that are Available ON or Available OFF. The lower and upper temperature bounds are given by $\underline{\theta}^k = \theta_r^k - \Delta^k$ and $\overline{\theta}^k = \theta_r^k + \Delta^k$. }\label{fig:priority_stacks}
\end{figure}

In practice, it is more favorable to turn ON (or OFF) the units which are going to be turned ON (or OFF) by their local hysteretic control law. To this end, we propose a priority-stack-based control method. The unit with the highest priority will be turned ON (or OFF) first, and then units with lower priorities will be considered in sequence until the desired regulation is achieved. This priority-stack-based control strategy minimizes the ON/OFF switching action for each unit, reducing wear and tear of \acp{TCL}. Priority stacks are illustrated in Fig.~\ref{fig:priority_stacks}.
A normalized {\em temperature distance} is considered as the sorting criterion.
%

\subsection{No-Short-Cycling Constraint}
The proposed priority-stack-based control scheme attempts to reduce the consecutive switching times of each \ac{TCL}. However, it can not guarantee that none of the units will not be switched quicker than allowed. To this end, one should \emph{explicitly} impose such no-short-cycling constraints on the priority stacks. As shown in Fig.~\ref{fig:priority_stacks}, the ON and OFF priority stacks can be modified to account for such no-short-cycling constraints.
Once a unit is turned ON or OFF, it must remain in that state for at least a certain amount of time (that is specified by the manufacture) before it is switched again.
\begin{table*}[tb]
\begin{center}
\begin{tabular}{cc}
\hline
Term & Description \\
\hline
$P^{k}_m$  & Power draw of unit $k$ when ON\\
$P_{tot}$ & $\sum_{k} P^{k}_m$\\
$P_{o}^{k}$ & Average power draw of unit $k$\\
$P_{ave}$ & $\sum_{k} P_{o}^{k}$\\
$r(t)$ & Regulation signal requesting power draw of $P_{ave} + r(t)$\\
Available ON & Units that have been ON for more than a certain amount of time\\
Unavailable ON & Units that have been ON for less than a certain amount of time\\
Available OFF & Units that have been OFF for more than a certain amount of time\\
Unavailable OFF & Units that have been OFF for less than a certain amount of time\\
$P_{ON\rightarrow OFF}^{lim}(t)$ & Total power of units switched from ON to OFF at  time $t$ due to temperature bound\\
$P_{OFF\rightarrow ON}^{lim}(t)$ & Total power of units switched from OFF to ON at  time $t$ due to temperature bound\\
$P_{ON}(t)$ & Total power of ON units\\
$P_{OFF}(t)$ & Total power of OFF units\\
$P_{ON}^{avail}(t)$  & Total power of units that are available ON\\
$P_{ON}^{unavail}(t)$  & Total power of units that are unavailable ON\\
$P_{OFF}^{avail}(t)$ & Total power of units that are available OFF\\ 
$P_{OFF}^{unavail}(t)$ & Total power of units that are unavailable OFF\\ 
\hline
\end{tabular}
\end{center}
\caption{Nomenclature of some of the frequently-used terms.}
\label{tab:nomenclature}
\end{table*}
For clarity of presentation, we list some of the terms that we will frequently use in this section in~Table~\ref{tab:nomenclature}.
When the controller must satisfy the no-short-cycling constraints, a certain percentage of \acp{TCL} will be unavailable to be switched from ON to OFF or OFF to ON. The effect of this loss of use is to create an additional constraint on {\em changes} in feasible regulation signals $r(t)$. Quite simply, if there is no available ON unit to be switched OFF, the regulation signal cannot request decreased power draw (and similarly for increased power draw). To determine feasible regulation signals, the battery model must be augmented with the constraints
\begin{equation}
- \mu_{-}(t) \leq \Delta r(t) \leq \mu_{+}(t), 
\label{eqn:deltafeasible}
\end{equation}
where $\Delta r(t) = r(t)-r(t-1)$, and $\mu_{-}(t)$ and $\mu_{+}(t)$ are time varying constraints.\footnote{There might exist other known constraints/bounds on changes $\Delta r(t)$ specified by the system operator which we are not considering here.} However, we will show that $\mu_{-}$ and $\mu_{+}$ are easily estimated if the following information is available: (i) power draw of each unit (when ON); (ii) average power draw of each unit $P_o^k$; and (iii) the total rated power for units that are about to be turned ON or OFF due to their temperature limits, denoted by $P_{OFF\rightarrow ON}^{lim}(t)$ or $P_{ON\rightarrow OFF}^{lim}(t)$, respectively.  
%
%
\begin{theorem}
Assume a collection of \acp{TCL} defined by $P_m^{k}, P_{o}^{k}$, and a minimum short cycle time of $\tau$ (samples). If the regulation signal $r(t)$ has been met through sample time $t$, then the total power of units available at $t$ is 
\begin{multline*}
P^{avail}_{OFF}(t) = P_{tot} - P_{ave} - r(t) - \\ \sum_{k=t-\tau}^{t} \left(P^{lim}_{ON\rightarrow OFF}(k) + \left[-D(k)\right]_{+}\right)
\end{multline*}
and
\begin{multline*}
\hspace{-0.1in}
P^{avail}_{ON}(t) = P_{ave} + r(t) - \sum_{k=t-\tau}^{t} \left( P^{lim}_{OFF\rightarrow ON}(k) +  \left[D(k)\right]_{+}\right),
\end{multline*}
where $D(t):=\Delta r(t) - (P_{OFF\rightarrow ON}^{lim}(t) - P_{ON\rightarrow OFF}^{lim}(t))$ and $[x]_{+} := \max(x,0)$. In addition, feasible $\Delta r(t)$ satisfies (\ref{eqn:deltafeasible}) with
\begin{align*}
\mu_{+}(t) = P_{OFF}^{avail}(t-1) - \max(P^{lim}_{ON\rightarrow OFF}(t),P^{lim}_{OFF\rightarrow ON}(t)),\\
\mu_{-}(t) = P_{ON}^{avail}(t-1) - \max(P^{lim}_{ON\rightarrow OFF}(t),P^{lim}_{OFF\rightarrow ON}(t)).
\end{align*}
\label{theo:ramping_rate_constraint}
\end{theorem}
\begin{proof}
See Appendix.
\end{proof}

\subsection{Simulation Results}
\begin{figure*}
\begin{center}
\includegraphics[width=0.95\columnwidth]{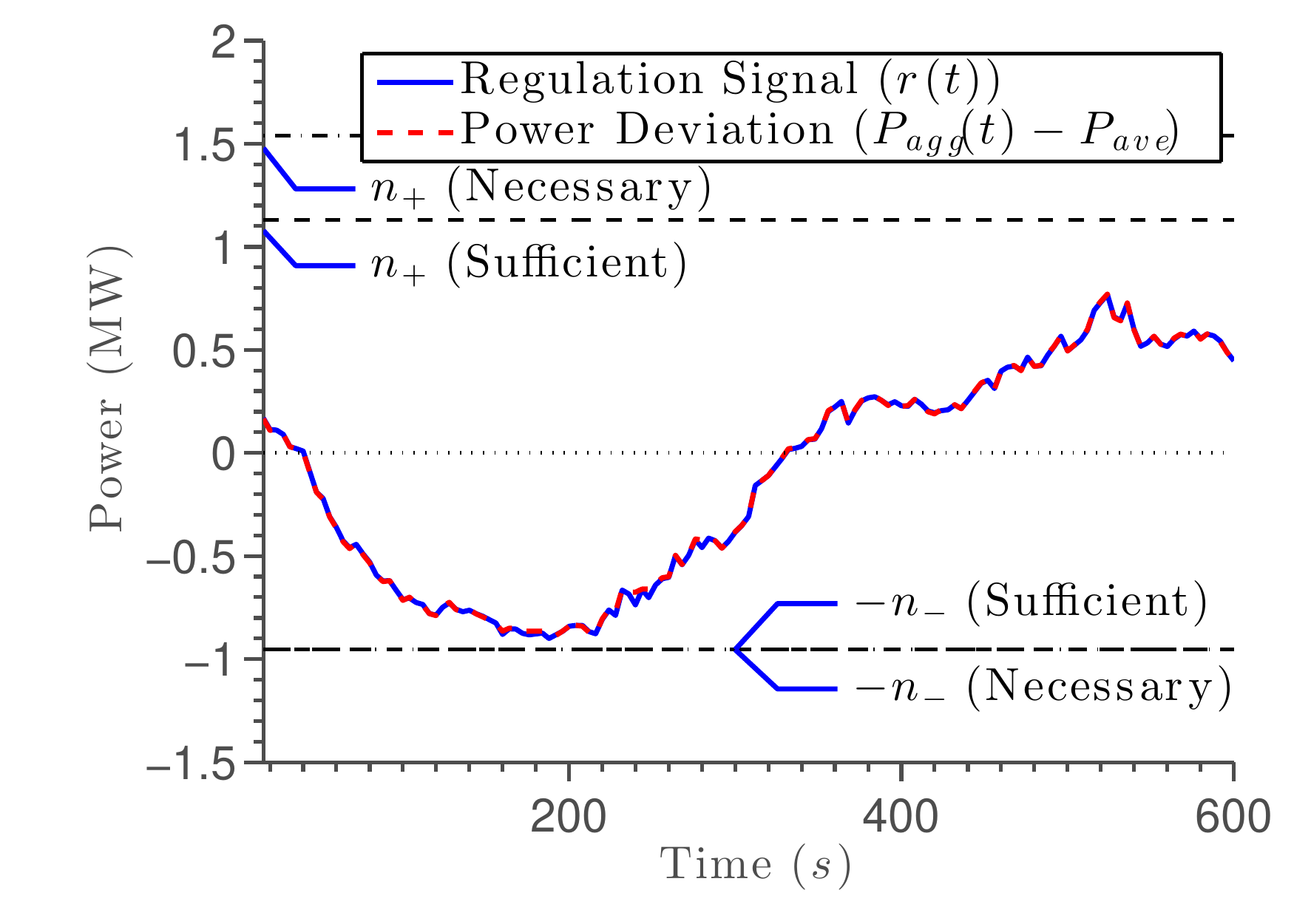} 
(a)
\includegraphics[width=0.95\columnwidth]{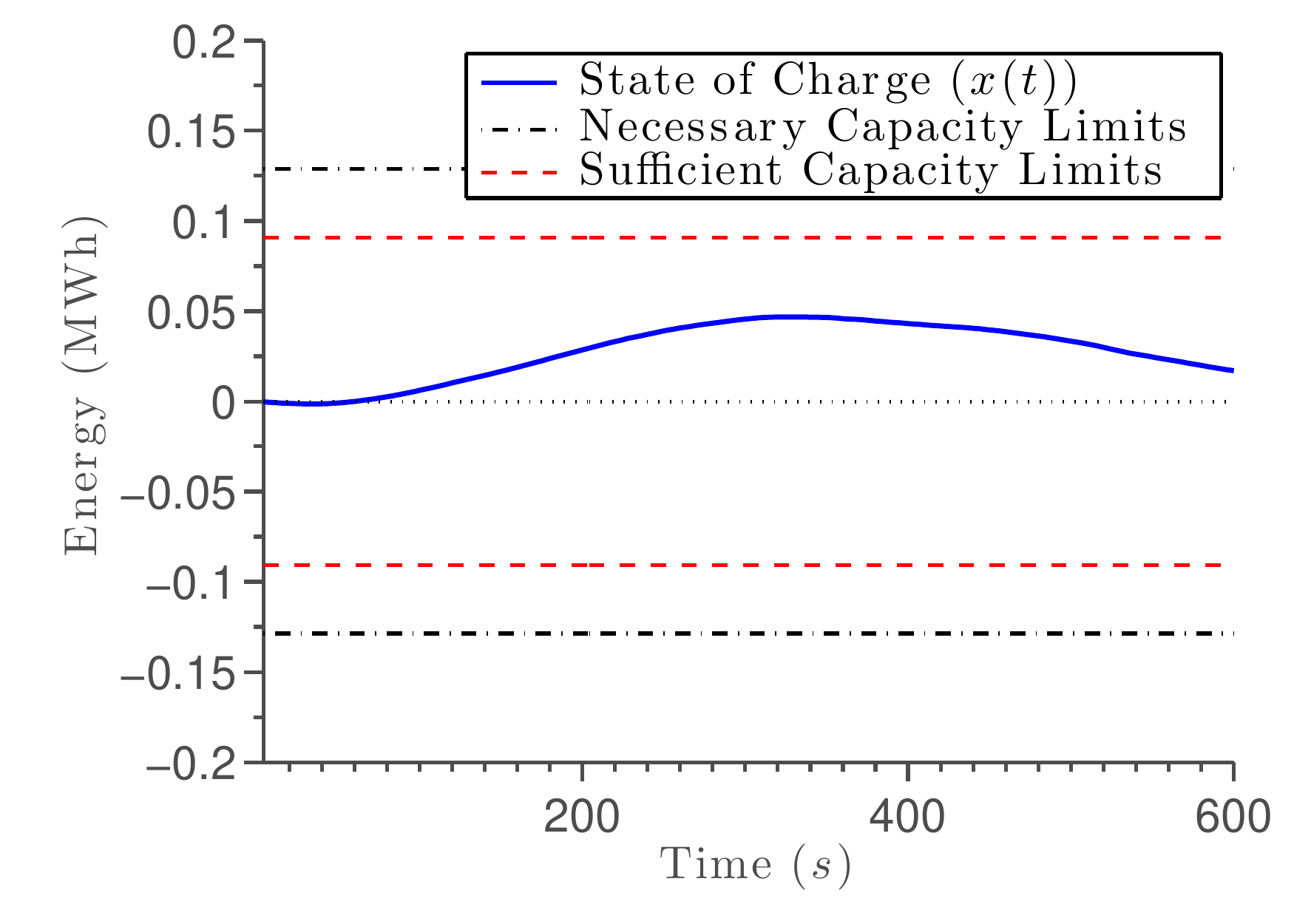} 
(b)\\
\includegraphics[width=0.95\columnwidth]{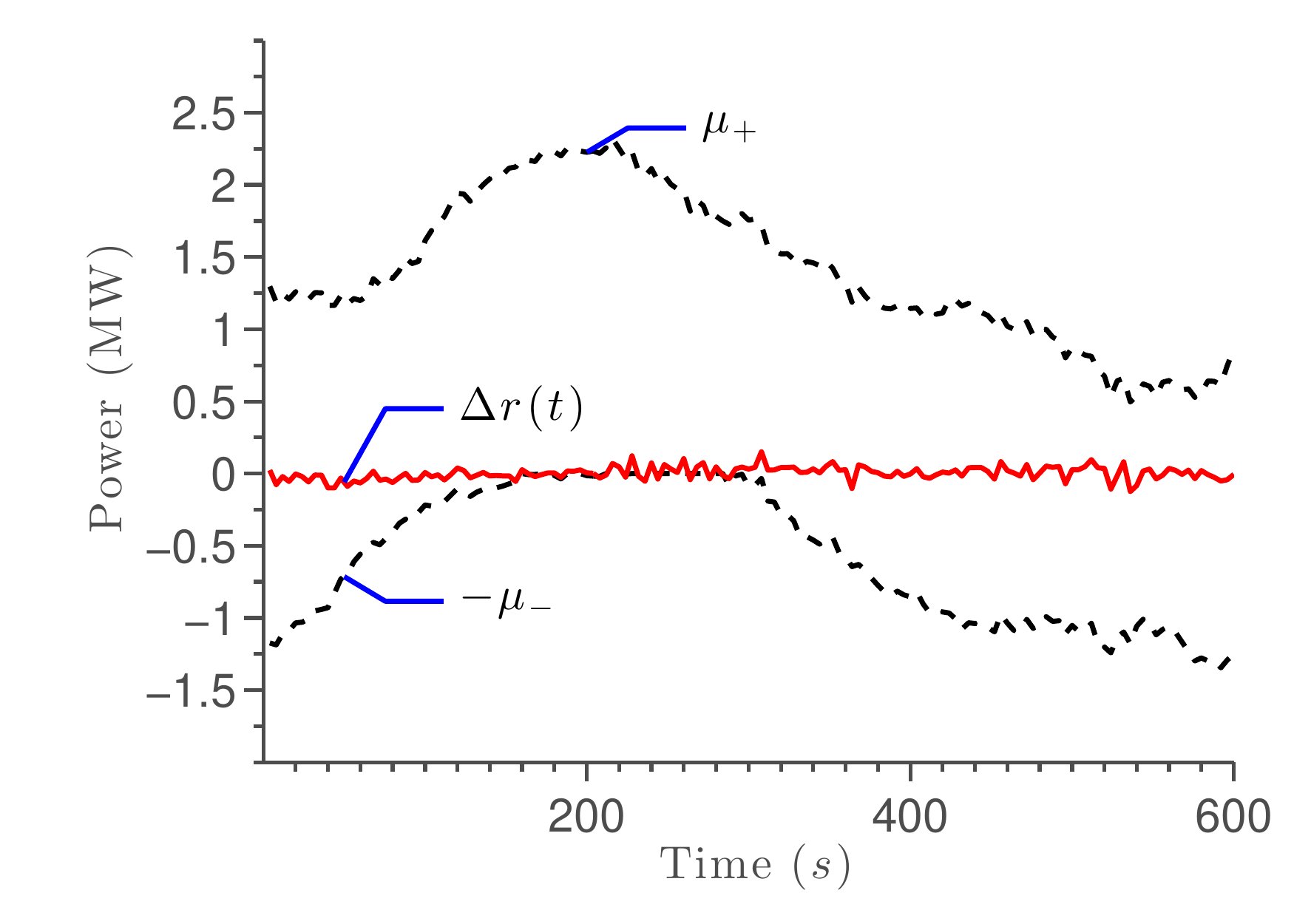} 
(c)
\includegraphics[width=0.95\columnwidth]{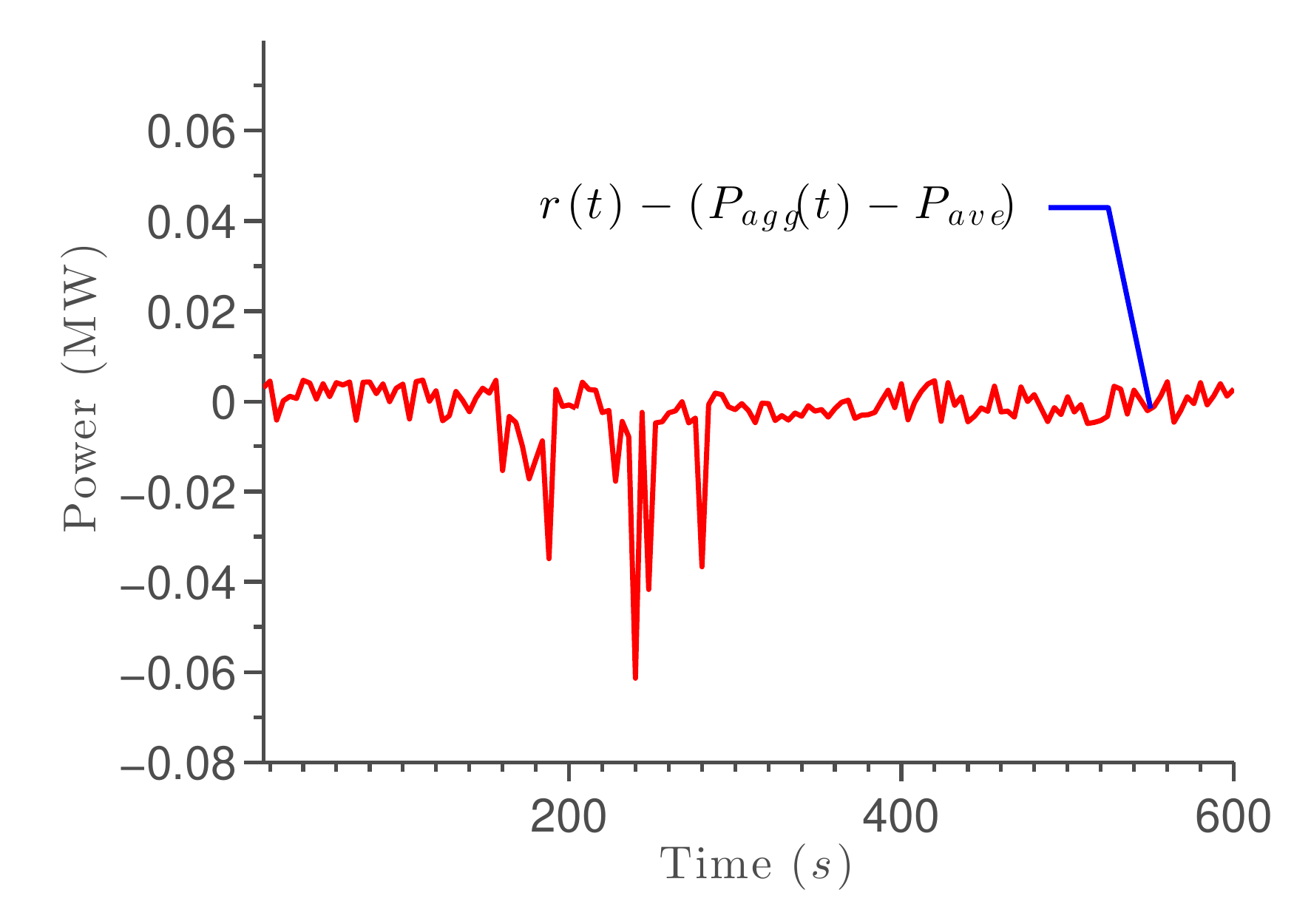} 
(d)
\end{center}
\caption{Illustration of the effect of short cycling and ramping rate constraints. (a) The regulation signal and battery model bounds on power. (b) The \ac{SoC} and the capacity limits. (c) $\Delta r(t)$ and its bounds given in Theorem~\ref{theo:ramping_rate_constraint}. (d) The difference between the desired regulation signal $r(t)$ and the actual power draw $P_{agg}(t) - P_{ave}$. At time $150$ (s), $-\mu_-(t)$ approaches zero, meaning that negative $\Delta r(t)$ is no longer feasible. Fig.~\ref{fig:shortcycle}(d) confirms that the regulation signal is not well followed downward during the time that $-\mu_{-}(t)$ is close to zero.}
\label{fig:shortcycle}
\end{figure*}

We run the priority-stack-based controller with the reference command shown as the solid line in Fig.~\ref{fig:shortcycle}(a). For comparison, the power and capacity limits found using the battery model are also shown in Fig.~\ref{fig:shortcycle}(a) and Fig.~\ref{fig:shortcycle}(b), respectively. As can be seen, the power and capacity limits are not violated by this regulation signal.
We take $ P^{lim}_{OFF\rightarrow ON}(t)$ and $ P^{lim}_{OFF\rightarrow ON}(t)$ as that reported by the local unit controllers, and use that information along with $r(t)$ to calculate $\mu_{+}(t) $ and $\mu_{-}(t)$. In Fig.~\ref{fig:shortcycle}(c) these are plotted along with $\Delta r(t)$. Note that at time $150$ (s), the lower bound approaches zero, meaning that negative $\Delta r(t)$ is no longer feasible. Fig.~\ref{fig:shortcycle}(d) depicts the difference between the desired regulation signal and the actual power draw $P_{agg}(t)-P_{ave}$, confirming that regulation signal is not well followed downward during the time that $\mu_{-}$ is close or equal to zero. 

%
\section{Conclusions and Future Work}
\label{sec:conclusion}
An aggregation of Thermostatically Controlled Loads (TCLs) can be utilized to provide fast regulating reserve service for power grids and the behavior of the aggregation can be captured by a stochastic battery with dissipation.
In this paper, we addressed two practical issues associated with the proposed battery model. First, we addressed clustering of a heterogeneous collection of TCLs and showed that by finding the optimal dissipation parameter for a given collection, one can divide these units into few clusters and improve the overall battery model.
Second, we analytically characterized the impact of imposing a no-short-cycling requirement on TCLs in terms of constraints on the ramping rate of the AGC signal. 

One of the future directions of this work is to better understand the thermal characteristics and dynamics of an individual \ac{TCL}. We should also better understand the actual heterogeneity pattern of a collection of \acp{TCL}. These can be done by examining the proposed models against experimental data captured from installed \acp{TCL}.
   

\appendix
{\textbf{Proof of Lemma~\ref{lem:hetero_thermal}}}
Let $C_{\min} := \min_k C^k$ and $C_{\max} := \max_k C^k$. If $\alpha > 1/RC_{\min}$, then $\alpha RC^k>1$, $\forall k$, and $g^k = 1/\alpha R$. If $\alpha<1/RC_{\max}$, then $\alpha RC^k<1$, $\forall k$, and $g^k = \frac{C^k}{2-\alpha RC^k}$. With a change of variable $x := 1/RC^k$, $g^k(x) = 1/(2Rx-\alpha R)$.
If $1/RC_{\max} < \alpha < 1/RC_{\min}$, for $C^k$'s such that $C^k>1/\alpha R$, $g^k = 1/\alpha R$, and for $C^k$'s such that $C^k<1/\alpha R$, $g^k = \frac{C^k}{2-\alpha RC^k}$. 
%
At $x = \alpha$, $1/(2Rx-\alpha R) = 1/\alpha R$. 
Consequently, when $\alpha<1/RC_{\min}$, $\min_k g^k = 1/(2/C_{\min} - \alpha R)$ and when $\alpha>1/RC_{\min}$, $\min_k g^k= 1/\alpha R$.
%
Thus, when the heterogeneity is only in $C^k$,
\[
\max_{\alpha} \mathcal{C}(\alpha) = N\Delta C_{\min}/\eta, \qquad \alpha^* = 1/RC_{\min}.
\]


{\textbf{Proof of Lemma~\ref{lem:hetero_deadband}}}
If $\alpha > 1/RC$, then $\mathcal{C}(\alpha)= N\Delta_{\min}/ \alpha \eta R$. If $\alpha<1/RC$, then $\mathcal{C}(\alpha) = N\Delta_{\min}C/\eta(2-\alpha RC)$. 
The breakpoint is at $\alpha = 1/RC$. Thus, when the heterogeneity is only in $\Delta^k$,
\[
\max_{\alpha} \mathcal{C}(\alpha) = NC\Delta_{\min}/\eta \ \ \text{and} \ \ \alpha^* = 1/RC.
\]
%


{\textbf{Proof of Theorem~\ref{theo:clustering}}}
When the heterogeneity is only in $C^k$, the optimal cluster sizes can be found by solving the following optimization problem:
\begin{equation}
\begin{aligned}
& \underset{N_1,\dots,N_m}{\text{maximize}}
& & f(1)N_1 + \sum_{i=2}^mN_if\big(1+\sum_{j=1}^{i-1}N_j\big)\\
& \text{subject to}
& & \sum_{i=1}^m N_i = N,
\end{aligned}
\label{eq:max_cluster_generic_1}
\end{equation}
where $f(\cdot)$ is a function that represents the sorted $C^k$ values in an ascending order.
%
In the case where a uniform distribution is assumed as the heterogeneity of $C^k$'s, the sorted $C^k$ values construct an affine function $f$ between $C_{\min}$ and $C_{\max}$ as
\[
f(x) = C_{\min} + \frac{C_{\max}-C_{\min}}{N-1}(x-1),
\]
where $x$ only takes integer values between $1$ and $N$. It can be shown that under linearity assumption on $f(\cdot)$, the optimal solution to (\ref{eq:max_cluster_generic_1}) is 
\[
N_1^* = \dots = N_m^* = N/m. 
\] 
The proof is not presented here for the sake of saving space. Consequently, the optimal capacity is derived by using the optimal cluster sizes in the objective function of~(\ref{eq:max_cluster_generic_1}) as
\[
\mathcal{C}^*_m = \big(C_{\min}+\frac{(C_{\max}-C_{\min})}{2}\frac{N}{(N-1)}\frac{(m-1)}{m}\big)N\Delta/\eta.
\]

{\textbf{Proof of Theorem~\ref{theo:ramping_rate_constraint}}}
Let $P_{ON}(t)$ and $P_{OFF}(t)$ denote the total power of units ON and OFF, respectively. If $r(t)$ is satisfied, then by definition
\begin{align*}
P_{ON}(t) & = P_{ave} + r(t), \\
P_{OFF}(t) & = P_{tot} - P_{ave} - r(t).
\end{align*}
Note that $P_{ON}(t) + P_{OFF}(t) = P_{tot}$. Let $P_{ON}^{unavail}(t)$ and $P_{OFF}^{unavail}(t)$ be the total power of units that are unavailable and ON or OFF, respectively. Clearly
\begin{align*}
P^{avail}_{ON}(t) & = P_{ON}(t)  - P_{ON}^{unavail}(t),\\
P^{avail}_{OFF}(t) & = P_{OFF}(t)  - P_{OFF}^{unavail}(t).
\end{align*} 
Now, $P_{ON}^{unavail}$ is given by the sum of the power of units that have been turned ON in the last $\tau$ seconds. The first result follows by noting that if $r(t)$ is satisfied, then the power of units turned ON at time $t$ must balance the difference between units turned ON and OFF due to local controllers, along with the change in $r(t)$. For example, the units turned from OFF to ON must be given by
\[
P_{OFF\rightarrow ON} (t)=P^{lim}_{OFF\rightarrow ON}(t) + \left[D(k)\right]_{+},
\]
where the second term represents the potential imbalance due to a change in $\Delta r(t)$ plus a difference between $P_{OFF\rightarrow ON}^{lim}(t)$ and $P_{ON\rightarrow OFF}^{lim}(t)$. Finally, the limits $\mu_{+}(t)$ and $\mu_{-}(t)$ are achieved using the worst case assumption that all units that hit the temperature limits are currently unavailable and no units unavailable at time $t-1$ become available.


\bibliographystyle{IEEEtran}    
\bibliography{ACC14-TCL-Clustering}

\begin{thebibliography}{10}
\providecommand{\url}[1]{#1}
\csname url@rmstyle\endcsname
\providecommand{\newblock}{\relax}
\providecommand{\bibinfo}[2]{#2}
\providecommand\BIBentrySTDinterwordspacing{\spaceskip=0pt\relax}
\providecommand\BIBentryALTinterwordstretchfactor{4}
\providecommand\BIBentryALTinterwordspacing{\spaceskip=\fontdimen2\font plus
\BIBentryALTinterwordstretchfactor\fontdimen3\font minus
  \fontdimen4\font\relax}
\providecommand\BIBforeignlanguage[2]{{%
\expandafter\ifx\csname l@#1\endcsname\relax
\typeout{** WARNING: IEEEtran.bst: No hyphenation pattern has been}%
\typeout{** loaded for the language `#1'. Using the pattern for}%
\typeout{** the default language instead.}%
\else
\language=\csname l@#1\endcsname
\fi
#2}}

\bibitem{CA_renewable_portfolio}
\BIBentryALTinterwordspacing
{California Energy Commission}, ``California renewable energy overview and
  programs,'' 2013. [Online]. Available:
  \url{http://www.energy.ca.gov/renewables/index.html}
\BIBentrySTDinterwordspacing

\bibitem{smith2007utility}
J.~C. Smith, M.~R. Milligan, E.~A. DeMeo, and B.~Parsons, ``Utility wind
  integration and operating impact state of the art,'' \emph{IEEE Transactions
  on Power Systems}, vol.~22, no.~3, pp. 900 --908, 2007.

\bibitem{makarov2009operational}
Y.~V. Makarov, C.~Loutan, J.~Ma, and P.~de~Mello, ``Operational impacts of wind
  generation on california power systems,'' \emph{IEEE Transactions on Power
  Systems}, vol.~24, no.~2, pp. 1039 --1050, 2009.

\bibitem{meynegwankowsha10}
S.~Meyn, M.~Negrete-Pincetic, G.~Wang, A.~Kowli, and E.~Shafieepoorfard, ``The
  value of volatile resources in electricity markets,'' in \emph{CDC2010},
  2010, pp. 1029 --1036, and submitted to {IEEE TAC}, 2012.

\bibitem{AS_Kirby}
B.~Kirby, ``Ancillary services: {T}echnical and commercial insights,'' Report
  prepared for Wartsila, Tech. Rep., July 2007.

\bibitem{helman2010resource}
U.~Helman, ``Resource and transmission planning to achieve a 33\% {RPS} in
  {California}--{ISO} modeling tools and planning framework,'' in \emph{FERC
  Technical Conference on Planning Models and Software}, 2010.

\bibitem{CAISO_flexible}
\BIBentryALTinterwordspacing
{Market and Infrastructure Policy}, ``2013 flexible capacity procurement
  requirement,'' {California Independent System Operator -- CAISO}, Tech. Rep.,
  March 2012. [Online]. Available: \url{http://www.CAISO.com/}
\BIBentrySTDinterwordspacing

\bibitem{kema2009fastresponse}
K.~Vu, R.~Masiello, and R.~Fioravanti, ``Benefits of fast-response storage
  devices for system regulation in {ISO} markets,'' \emph{IEEE Power Energy
  Society General Meeting, 2009}, pp. 1 --8, 2009.

\bibitem{pnnl2008value}
Y.~V. Makarov, L.~S., J.~Ma, and T.~B. Nguyen, ``Assessing the value of
  regulation resources based on their time response characteristics,'' Pacific
  Northwest National Laboratory, Tech. Rep. PNNL-17632, 2008.

\bibitem{ferc_755}
\BIBentryALTinterwordspacing
{Market and Infrastructure Policy}, ``Frequency regulation compensation -
  {FERC} order no. 755,'' {Federal Energy Regulatory Commission -- FERC}, Tech.
  Rep., 2011. [Online]. Available:
  \url{https://www.ferc.gov/whats-new/comm-meet/2011/102011/E-28.pdf}
\BIBentrySTDinterwordspacing

\bibitem{ferc_784}
\BIBentryALTinterwordspacing
------, ``Frequency regulation compensation - {FERC} order no. 784,'' {Federal
  Energy Regulatory Commission -- FERC}, Tech. Rep., 2013. [Online]. Available:
  \url{https://www.ferc.gov/whats-new/comm-meet/2013/071813/E-22.pdf}
\BIBentrySTDinterwordspacing

\bibitem{callaway2009tapping}
D.~S. Callaway, ``Tapping the energy storage potential in electric loads to
  deliver load following and regulation, with application to wind energy,''
  \emph{Energy Conversion and Management}, vol.~50, no.~5, pp. 1389--1400,
  2009.

\bibitem{koch2009active}
S.~Koch, M.~Zima, and G.~Andersson, ``Active coordination of thermal household
  appliances for load management purposes,'' \emph{Proceedings of the IFAC
  Symposium on Power Plants and Power Systems Control}, pp. 149--154, 2009.

\bibitem{koch2011modeling}
S.~Koch, J.~Mathieu, and D.~Callaway, ``Modeling and control of aggregated
  heterogeneous thermostatically controlled loads for ancillary services,''
  \emph{Proceedings of the $17$th Power Systems Computation Conference --
  PSCC}, 2011.

\bibitem{mathieu2012state}
J.~Mathieu and D.~Callaway, ``State estimation and control of heterogeneous
  thermostatically controlled loads for load following,'' \emph{Proceedings of
  the $45$th Hawaii International Conference on System Sciences -- HICCS45},
  pp. 2002--2011, 2012.

\bibitem{mathieu2013state}
J.~L. Mathieu, S.~Koch, and D.~S. Callaway, ``State estimation and control of
  electric loads to manage real-time energy imbalance,'' \emph{IEEE
  Transactions on Power Systems}, vol.~28, no.~1, pp. 430 --440, Feburary 2013.

\bibitem{mathieu2013energy}
J.~L. Mathieu, M.~Kamgarpour, J.~Lygeros, and D.~S. Callaway, ``Energy
  arbitrage with thermostatically controlled loads,'' in \emph{European Control
  Conference (ECC)}, 2013.

\bibitem{zhang2012aggregate}
W.~Zhang, K.~Kalsi, J.~Fuller, M.~Elizondo, , and D.~Chassin, ``Aggregate model
  for heterogeneous thermostatically controlled loads with demand response,''
  \emph{Proceedings of the 2012 IEEE PES General Meeting}, 2012.

\bibitem{chang2013modeling}
C.-Y. Chang, W.~Zhang, J.~Lian, and K.~Kalsi, ``Modeling and control of
  aggregated air conditioning loads under realistic conditions,''
  \emph{Proceedings of the IEEE PES Innovative Smart Grid Technologies
  Conference -- ISGT}, 2013.

\bibitem{zhang2013aggregated}
W.~Zhang, J.~Lian, C.-Y. Chang, and K.~Kalsi, ``Aggregated modeling and control
  of air conditioning loads for demand response,'' \emph{IEEE Transactions on
  Power Systems}, vol.~28, no.~4, pp. 4655 -- 4664, 2013.

\bibitem{bashash2011modeling}
S.~Bashash and H.~K. Fathy, ``Modeling and control insights into demand-side
  energy management through setpoint control of thermostatic loads,''
  \emph{Proceedings of the 2011 American Control Conference -- ACC}, pp.
  4546--4553, 2011.

\bibitem{bashash2013modeling}
------, ``Modeling and control of aggregate air conditioning loads for robust
  renewable power management,'' \emph{IEEE Transactions on Control Systems
  Technology}, vol.~21, no.~4, pp. 1318--1327, 2013.

\bibitem{sanandaji2014fast}
B.~M. Sanandaji, H.~Hao, and K.~Poolla, ``Fast regulation service provision via
  aggregation of thermostatically controlled loads,'' \emph{Proceedings of the
  $47$th Hawaii International Conference on System Sciences -- HICSS47}, pp.
  2388--2397, 2014.

\bibitem{bashash2011robust}
S.~Bashash and H.~K. Fathy, ``Robust demand-side plug-in electric vehicle load
  control for renewable energy management,'' \emph{Proceedings of the 2011
  American Control Conference -- ACC}, pp. 929--934, 2011.

\bibitem{nayyar2013EV}
A.~Nayyar, J.~Taylor, A.~Subramanian, K.~Poolla, and P.~Varaiya, ``Aggregate
  flexibility of a collection loads,'' \emph{Proceedings of the $52$th IEEE
  Conference on Decision and Control -- CDC}, 2013.

\bibitem{maasoumy2014flexibility}
M.~Maasoumy, B.~M. Sanandaji, K.~Poolla, and A.~Sangiovanni-Vincentelli,
  ``Model predictive control of regulation services from commercial buildings
  to the smart grid,'' \emph{Proceedings of the 2014 American Control
  Conference -- ACC}, 2014.

\bibitem{building_energy_data_book}
\BIBentryALTinterwordspacing
``Buildings energy data book.'' [Online]. Available:
  \url{http://buildingsdatabook.eren.doe.gov/default.aspx}
\BIBentrySTDinterwordspacing

\bibitem{eia_aer}
\BIBentryALTinterwordspacing
``{U.S. Energy Information Administration}, annual energy review,'' 2010.
  [Online]. Available:
  \url{http://www.eia.gov/totalenergy/data/annual/#consumption}
\BIBentrySTDinterwordspacing

\bibitem{kundu2011modeling}
S.~Kundu, N.~Sinitsyn, S.~Backhaus, and I.~Hiskens, ``Modeling and control of
  thermostatically controlled loads,'' \emph{Proceedings of the $17$th Power
  Systems Computation Conference -- PSCC}, 2011.

\bibitem{hehao2013generalized}
H.~Hao, B.~M. Sanandaji, K.~Poolla, and T.~L. Vincent, ``{A Generalized Battery
  Model of a Collection of Thermostatically Controlled Loads for Providing
  Ancillary Service},'' \emph{Proceedings of the $51$th Annual Allerton
  Conference on Communication, Control and Computing}, pp. 551--558, 2013.

\bibitem{hehao2013aggregate}
\BIBentryALTinterwordspacing
------, ``Aggregate flexibility of thermostatically controlled loads,''
  \emph{IEEE Transactions of Power Systems, under review}, 2013. [Online].
  Available:
  \url{http://www.eecs.berkeley.edu/~sanandaji/my_papers/TPS_TCL.pdf}
\BIBentrySTDinterwordspacing

\bibitem{malhame1985electric}
R.~Malhame and C.-Y. Chong, ``Electric load model synthesis by diffusion
  approximation of a high-order hybrid-state stochastic system,'' \emph{IEEE
  Transactions on Automatic Control}, vol.~30, no.~9, pp. 854--860, 1985.

\bibitem{hehao2014frequency}
H.~Hao, B.~M. Sanandaji, K.~Poolla, and T.~L. Vincent, ``{Frequency Regulation
  from Flexible Loads: {P}otential, Economics, and Implementation},''
  \emph{Proceedings of the 2014 American Control Conference -- ACC}, 2014.

\end{thebibliography}

\end{document}